\documentclass[aps,twocolumn,floats,showpacs]{revtex4}
\usepackage{epsfig}
\usepackage{graphicx}
\usepackage{amsmath}

\newcommand{\beq}{\begin{equation}}
\newcommand{\eeq}{\end{equation}}
\newcommand{\bea}{\begin{eqnarray}}
\newcommand{\eea}{\end{eqnarray}}

\begin{document}

\title{Quantum heat transfer in harmonic chains with self consistent reservoirs:
Exact numerical simulations}

\author{Malay Bandyopadhyay}
\author{Dvira Segal}
\affiliation{Chemical Physics Theory Group, Department of Chemistry
 University of Toronto, 80 St. George
street, Toronto, Ontario, M5S 3H6, Canada}
\date{\today}

\begin{abstract}
We describe a numerical scheme for exactly simulating the heat
current behavior in a quantum harmonic chain with self-consistent
reservoirs. Numerically-exact results are compared to classical
simulations and to the quantum behavior under the linear response
approximation. In the classical limit or for small temperature
biases our results coincide with previous calculations. At large
bias and for low temperatures the quantum dynamics of the system
fundamentally differs from the close-to-equilibrium behavior,
revealing in particular the effect of thermal rectification for
asymmetric chains. Since this effect is absent in the classical
analog of our model, we conclude that in the quantum model studied
here thermal rectification is a purely quantum phenomenon, rooted in
the quantum statistics.
\end{abstract}

\pacs{63.22.-m, 44.10.+i, 05.60.-k, 02.70.-c}

\maketitle


\section{Introduction}

Understanding the role of quantum effects in the thermal conduction
properties of interacting systems is a challenging task. While in
the classical regime molecular dynamic simulations provide a
flexible tool for including anharmonic interactions to all orders
\cite{Lepri-rev,Dhar-rev}, in the quantum limit treatments are
typically limited to particular parameter domains \cite{Tu}. Among
the methods developed for tracking the quantum behavior of
anharmonic systems we recall the non-equilibrium Green's function
technique which is perturbative in the nonlinear interaction
strength \cite{Green,Wang1} and
the master equation approach, 
which is limited to models with weak-system bath couplings and to
systems with few quantum states \cite{masterD,Teemo}. More recently,
mixed classical-quantum molecular dynamics simulation tools were
developed, valid at relatively high temperatures \cite{Wang}. It was
also demonstrated that a scheme based on the Born-Oppenheimer
principle could be constructed in the context of thermal conduction,
useful for studying the dynamics in the off-resonance regime
\cite{BO}. Lastly, exact quantum simulations of the heat current
characteristics can be performed for simplified models only
\cite{Hartree}.

\begin{figure}[htbp]
\vspace{2mm} \hspace{2mm} {\hbox{\epsfxsize=75mm
\epsffile{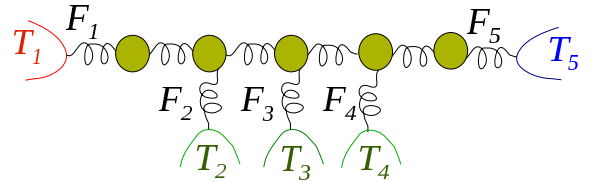}}} \vspace{3mm} \caption{Scheme of a harmonic
chain of $N=5$ beads, where the inner particles are connected to SC
baths. The wiggly lines represent harmonic bonds. The temperatures
$T_1$ and $T_5$ set the boundary conditions; the temperatures $T_l$
($l=2,3,4$) are determined by demanding that the leaking currents
vanish, $F_l=0$. The net heat current across the system is given by
$F_1=-F_5$.} \label{FigM}
\end{figure}

In this paper we consider the problem of heat transfer 
in the quantum harmonic chain model with each inner site connected
to a self consistent (SC) reservoir. This model has been developed
with the motivation to include nonlinear behavior in an effective
way \cite{Bolsterli,Rich,Bonetto,Dhar-rev}. Specifically, here we would
like to gain insight on the role of quantum effects at low
temperatures on the thermal properties of 1-dimensional (1D) chains.
For brevity, we often refer to this model as the ''SC model". It
includes a linear chain of $N$ beads connected to $N$ independent
thermal reservoirs, one at each site, see Fig. \ref{FigM}. While the
temperatures of the reservoirs attached to the first and last
particles impose the boundary conditions, the role of the inner
(self consistent) baths is to provide a simple scattering mechanism
that might lead to local equilibration and to the onset of the
(diffusional) Fourier's law of heat conduction \cite{Bonetto,
Dhar-rev}. In practice, the temperature of these $N-2$ internal
baths is determined by demanding that in steady-state, on average,
there is no net heat flow between the chain atoms and these
reservoirs.

The classical version of this model has been proposed in Refs.
\cite{Bolsterli,Rich} and recently revisited in \cite{Bonetto,PereiraC}, 
demonstrating that for long chains
Fourier's law is satisfied and a linear temperature profile is
generated. It has been also proved that in the
classical regime this model cannot support thermal rectification, an
asymmetry of the current under the exchange of the temperature bias,
even when some spatial asymmetry is provided \cite{Pereira,SegalSC}.
The SC model is also of interest in the context of anharmonic lattices
\cite{BonettoAn, PereiraAn}.
Overall, it is an example for a hybrid model, whose
time evolution is dictated both by a Hamiltonian term (deterministic), and by stochastic effects.


The quantum analog of the SC model was studied by Visscher and Rich
\cite{RichQ}, who analyzed the limiting case of weak-coupling to the
SC reservoirs. More recently, the model was revisited by Roy and
Dhar \cite{Roy1, Roy2}, who demonstrated that under the {\it linear
response assumption} and for asymptotically long chains, Fourier's
law holds and a temperature dependent thermal conductivity is
realized. An analytical study of the SC model with alternate masses
has revealed the role of quantum effects at low temperatures
\cite{Neto,PereiraQ}. More recently, a mathematical analysis of the
mass-graded SC model in the quantum domain has indicated on the
onset of thermal rectification, beyond the linear response regime
\cite{PereiraPLA}.

Focusing on the quantum SC harmonic chain model, one should note
that exact analytic results are limited to the linear response
regime \cite{Roy1,Roy2}. The reason is that beyond this limit, for
large temperature biases, the self consistent condition translates
into a set of coupled {\it nonlinear} equations which seem
intractable. Since an analytic solution is missing, in this paper we
suggest a numerical scheme for exactly simulating the transport
properties of this model. The method is useful at low and high
temperatures, in equilibrium, and for far-from equilibrium
situations. It can be also applied onto three-dimensional systems.
For simplicity, here we confine ourselves to 1D models.
%
As an interesting application we perform numerical simulations on
the SC harmonic chain model, incorporating a spatial asymmetry.
In accordance with previous analytic indications \cite{PereiraPLA},
we confirm that the system rectifies heat in the quantum regime,
beyond the linear response limit.

The paper is organized as follows. In Sec. II we present the SC
harmonic chain model and describe our numerical method. We further
explain how to calculate the thermal properties of the model in
the classical limit, and in the quantum regime, under the
linear-response approximation. Section III provides some examples
for the heat current characteristics in different domains,
manifesting the onset of the thermal rectifying effect in an
asymmetric setting. Section IV concludes.

\section{Model and Method}

We now describe the SC model, introduced in Ref. \cite{Rich}. The
chain includes $N$ atoms, where neighbors are connected by harmonic
links. Each particle is also bilinearly (position-position) coupled
to an independent thermal reservoir. The temperatures at the end
points are set to $T_1$ and $T_N$, establishing the boundary
conditions. In contrast, the temperatures of the internal reservoirs
$T_l$ ($l=2,3,...,N-1$) are determined in a self consistent manner, by
requiring that the net heat current flowing into or from the chain
through each contact $l=2,3,...,N-1$ vanishes. For a schematic
representation see Fig. \ref{FigM}.

The Hamiltonian of the chain $H_S$ (system),
its reservoirs $H_{B_l}$ (baths), and the interaction term $\mathcal V_{B_l}$
is given as the sum of quadratic terms,
\bea
H=H_S + \sum_{l=1}^N H_{B_l} +\sum_{l=1}^N \mathcal{V}_{B_l},
\label{eq:H}
\eea
where
\bea
H_S &=& \frac{1}{2}\dot{X}_S^TM_S\dot{X}_S+\frac{1}{2}X_S^T\Phi_S X_S, \nonumber \\
H_{B_l} &=& \frac{1}{2}\dot{X}_{B_l}^TM_B\dot{X}_{B_l}+\frac{1}{2}X_{B_l}^T\Phi_{B_l} X_{B_l}, \nonumber \\
\mathcal{V}_{B_l}&=& X_S^T V_{B_l} X_B.
\eea
Here $M_S$ and $M_B$ are real diagonal matrices representing the
masses of the particles in the chain and the bath particles,
respectively. The quadratic potential energies for the chain and
baths are given by the real symmetric matrices $\Phi_S$ and
$\Phi_{B_l}$, respectively. The term $V_{B_l}$ denotes the
interaction between the chain and the $l$th bath. The column vectors
$X_S$ and $X_{B_l}$ are the Heisenberg operators of the particle
displacements about some equilibrium configuration. In particular,
$X_S=\lbrace X_1, X_2,....,X_N\rbrace$ with $X_l$ as the position
operator of the $l$th particle of the chain. The momentum operators
are given by $\dot{X}=M^{-1}P$, where
$\lbrace X_l, P_l\rbrace$ satisfies the usual commutation relation,
$\lbrack X_l, P_m\rbrack=i\hbar \delta_{l,m}$.

Since the system is entirely harmonic, a generalized Langevin
equation for the system displacements can be written
\cite{heatcond,Dhar-rev,Roy1}. This is done by formally solving the
Heisenberg equations of motion (EOM) for the bath operators, then
plugging them into the EOM of the system displacements. In the ohmic
limit this results in
\bea
 M_l \ddot X_l=-(2X_l-X_{l-1}-X_{l+1})- \gamma_l\dot X_l +\eta_l.
\eea
Here $M_l$ is the mass of the $l$th particle and the force constants
are taken as unity. The chain-bath coupling strengths are enclosed
within the friction coefficients $\gamma_l$. The noise-noise
correlations, in frequency domain, satisfy
\bea
&&\frac{1}{2} \langle \eta_l(\omega)\eta_m(\omega')  +  \eta_l(\omega')\eta_m(\omega)\rangle
\nonumber\\
&&=\frac{\gamma_l \omega }{2\pi} \coth\left(\frac{\omega}{2T_l}\right)\delta(\omega+\omega')\delta_{l,m}.
\eea
%
The steady-state heat current can be obtained by evaluating (two-point)
position-momentum correlation functions \cite{Roy1}. Specifically, it can be shown that
the heat current from the $l$th reservoir into the chain is given by
\bea
F_l&=&\sum_{m=1}^{N}\gamma_l \gamma_m \int_{-\infty}^{\infty} d\omega \omega^2 | [G(\omega)]_{l,m} |^2
\frac{\omega}{\pi}
\nonumber\\
&\times&
\left[ f(\omega, T_l)- f(\omega, T_m)\right].
\label{eq:Fn0}
\eea
Here the matrix $G$ is the inverse of a tridiagonal matrix with
off-diagonal elements equal to -1 and diagonal elements $2
-M_l\omega^2-i\gamma_l\omega$, $f(\omega,T)=[e^{\omega/T}-1]^{-1}$
is the Bose-Einstein distribution. The temperature profile across
the system is obtained by {\it demanding} that
\bea
F_l=0, \,\,\,\,\,\,\, l=2,3,....,N-1.
\label{eq:Fn}
\eea
This condition translates into a set of $N-2$ nonlinear equations,
yielding the inner baths' temperatures $T_l$. Plugging the resulting
temperatures inside the expression for $F_1$ (or equivalently inside
$F_N$) yields the steady-state net heat current flowing across the
system,
\bea
J=F_1=-F_N.
\eea
An exact-analytic solution of Eq. (\ref{eq:Fn}) is generally not
accessible. A mathematical analysis has been carried out only
perturbatively for a specific model with $N=3$ \cite{PereiraPLA}. In
the linear response regime (or in the classical limit) these $N-2$
equations reduce into a set of {\it linear} equations, which can be
readily solved as we explain below. Here, with the motivation to
treat quantum systems beyond linear response, we develop an
iterative numerical scheme for acquiring the exact solution of Eq.
(\ref{eq:Fn}), i.e. the set of the inner-baths temperatures.


\begin{figure}[htbp]
\vspace{2mm}
{\hbox{\epsfxsize=80mm \epsffile{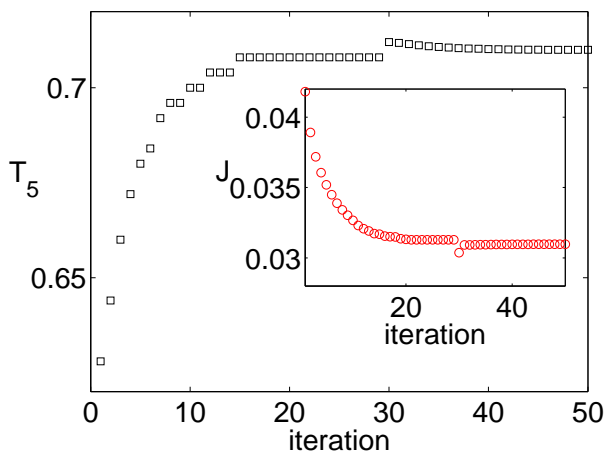}}} \vspace{3mm} \caption{
Convergence of the inner reservoirs' temperatures with increasing
number of iterations. $N=10$, $\beta_1=1$, $\beta_N=5$ and
$\gamma_n=0.2$ for $n=1,...,N$. The main plot shows the temperature
at site number 5. The inset displays the corresponding convergence
of the net heat current, $F_1$.} \label{Fig00}
\end{figure}

\begin{figure}[htbp]
\vspace{2mm}
{\hbox{\epsfxsize=80mm \epsffile{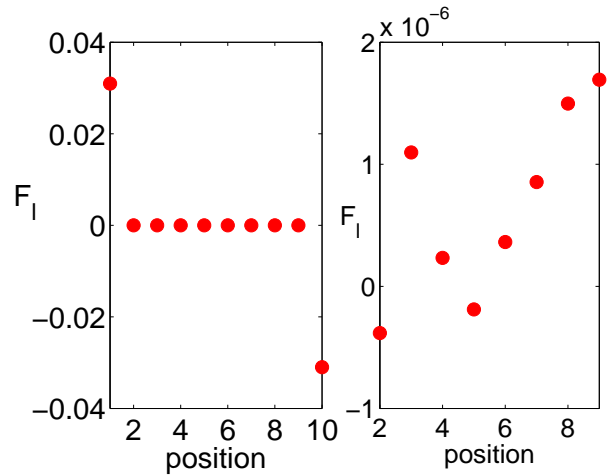}}}
\vspace{3mm}
\caption{
(left) The currents $F_l$, $ l=1,2...,N$ [see Eq. (\ref{eq:Fn0})],
at each site, for a chain with $N=10$ beads, obtained after applying the iterative procedure
 $p+q=50$ times,
$\beta_1=1$, $\beta_N=5$, $\gamma_l=0.2$. (right) Zooming on the
internal currents $F_2$ to $F_9$, zero for perfect SC reservoirs. }
\label{Fig0}
\end{figure}

\subsection{Quantum case: Exact simulations}

The nonlinear equations (\ref{eq:Fn}) can be numerically handled by
rearranging the expression for $F_l$ [Eq. (\ref{eq:Fn0})] as follows
\bea
\sum_{m=1}^NS_{l,m}(T_l)=\sum_{m=1}^NS_{l,m}(T_m),
 \label{eq:S}
\eea
where
\bea
S_{l,m}(T)= \gamma_l \gamma_m \int_{-\infty}^{\infty} d\omega \omega^2 | [G(\omega)]_{l,m} |^2
\frac{\omega}{\pi}  f(\omega, T).
\label{eq:Al}
\eea
Given $T_1$ and $T_N$, our goal is to obtain the temperatures of the
SC baths $T_l$ ($l=2,...,N-1$). This can be done by following an
iterative procedure. We first make an initial guess for the
temperature profile $T_l^{(0)}$. For example, we pick the average
temperature $T_l^{(0)}=(T_1+T_N)/2$. These values are inserted into
the {\it right hand side} of Eq. (\ref{eq:S}). For each site, we
then search for the value $T_l^{(1)}$
which yields an equality. 
We do it by calculating the {\it left hand side} of Eq. (\ref{eq:S}) over a fine grid of temperatures,
 searching for the temperature $T_l^{(1)}$ which minimizes the difference
$|\sum_{m=1}^NS_{l,m}(T_l^{(1)})-\sum_{m=1}^NS_{l,m}(T_m^{(0)})|$.
This process is repeated for each of the inner atoms, to obtain the
set of corrected temperatures $T_l^{(1)}$. In the next iteration
these temperatures are used as the basis supposition, for receiving
the subsequent corrected profile $T_l^{(2)}$. Formally, at the $k$th
step we solve the following equation $N-2$ times, for each inner
site $l$,
\bea
\sum_{m=1}^NS_{l,m}(T_l^{(k+1)})=\sum_{m=1}^NS_{l,m}(T_m^{(k)}).
\label{eq:it}
\eea
The procedure is repeated until we converge the temperature profile
$T_l$ and the current $F_1$. In other words, we maintain their
values through iterations. One should further verify that the
currents $F_l$, ($l=2,...,N-1$), flowing between the inner sites and
the SC reservoirs, are negligible, as we explain next.

There are two main sources of error in our procedure: (i) The
frequency integration in Eq. (\ref{eq:Al}) is carried out
numerically, by discretizing energies between a lower and upper
cutoffs. Selecting a fine frequency step $\Delta \omega$ and a large
energy cutoff $\omega_c\gg \omega_S$, with $\omega_S$ as the chain
characteristic frequency, we have verified that our results are
robust against $\Delta \omega$ and $\omega_c$. (ii) Equation
(\ref{eq:it}) is solved on a discretized temperature grid.
It is obvious that for a coarse grid the inner reservoirs'
temperatures may significantly deviate from the exact SC values and
leakage occurs \cite{leak}.
To control this  error we choose a mesh fine enough such that
$|F_l|/|F_1|<10^{-4}$ for $l=2,3,...,N-1$.
In addition, since for long chains the overall-net energy exchange between
the SC baths and the system may accumulate to large values, we also verify in our
simulations that the incoming and outgoing fluxes, $|F_1|$ and
$|F_N|$ (equal in principle) differ by less than 0.1$\%$.

In practice, we found that very delicate grids should be adopted for
reaching a good accuracy for chains with $N\gtrsim 10$. We have
therefore developed a two-step procedure to improve efficiency. In
the first step a relatively rough grid is constructed, $\delta
T=(T_1-T_N)/200$, and the iterative procedure [Eq. (\ref{eq:it})] is
followed to convergence, in the sense that the temperature profile
stays fixed through iterations. However, these temperatures still
deviate from the optimal (SC) temperatures, and significant leakage
takes place. We denote by $p$ the number of iterations in this part.

In the second step of our procedure an {\it individual} mesh is
constructed at each site by dividing the sector $[T_l^{(p)}-\delta
T, T_l^{(p)}+\delta T]$ into, say $200$ elements. Given these
individualized grids, we iterate Eq. (\ref{eq:it}) $q$ more times,
to converge the temperature profile again. Overall, $p+q$ iterations
are therefore performed, adopting a two-level grid. More generally,
one could use a hierarchy of temperature grids, individually
constructed around each particle for further improving the accuracy of the results
and the method efficiency.

Figures \ref{Fig00} and \ref{Fig0} demonstrate the convergence of
our scheme, as reflected in three quantities: (i) The temperature of
each internal bath should not vary between iterations upon convergence. In
conjunction, (ii) the current flowing through the system ($F_1$)
should remain fixed. (iii) Net exchange of energy between the internal
reservoirs and the chain should be rudimentary, relative to the
current crossing the system. It is important to note that one can
reach convergence with respect to the first two criteria, yet the
reservoirs may not act as SC baths since the temperature grid
selected is too rough.

For testing our method, we consider a uniform chain with $N=10$
particles of unit mass. The friction is uniform along the chain,
$\gamma_n=0.2$, $n=1,2,..,N$. Fig. \ref{Fig00} presents the
temperature at the chain center ($T_5$) as we repeatedly solve Eq.
(\ref{eq:it}). The small jump after 30 iterations arises due to the
re-definition (and refinement) of the temperature grid at this
point. The inset shows the (concurrent) convergence of the heat
current $F_1=-F_N$.
%
We also verify that the reservoirs indeed behave as SC baths. Fig.
\ref{Fig0} displays the currents in the system, in particular the
leakage currents $F_{l\neq 1,N}$ after 50 iterations. We confirm
that the local leakage  is smaller
by four orders of magnitude than the net heat current $F_1$ (right
panel). This assures us that at the end of the iterative procedure
the reservoirs serve as SC baths. 


\subsection{Quantum linear-response regime and classical calculations}

We outline here the process for obtaining the heat current behavior
for the SC model in the quantum linear response regime or in the
classical domain. In both cases, equation (\ref{eq:Fn}) reduces into
a set of linear equations.

In the quantum regime under the linear response approximation one
assumes that temperature differences along the chain are small,
$T_l-T_{l-1}\ll T_l$, thus the differences of Bose Einstein
functions ($f_l-f_m$) in Eq. (\ref{eq:Fn0}) can be replaced by the
derivative $ (T_l-T_m) \times \partial f/\partial T_a$ with
$T_a=(T_1+T_N)/2$. This approximation is valid close to equilibrium,
for $|T_1-T_N|\ll T_1, T_N$, or for very long chains with small
local gradients. Under this approximation Eq. (\ref{eq:Fn0}) reduces
to
\bea
F_l&=&\sum_{m=1}^{N}\gamma_l \gamma_m \int_{-\infty}^{\infty} d\omega \omega^2 | [G(\omega)]_{l,m} |^2
\frac{\omega}{\pi}
\nonumber\\
&\times& \frac{\omega}{4T_a^2} {\rm
csch}^2\left(\frac{\omega}{2T_a}\right) (T_l-T_m).
\label{eq:FnLR}
\eea
Similarly, in the classical limit the quantum statistics is replaced
by its high-temperature limit, $f(\omega,T)\sim T/\omega$ and Eq.
(\ref{eq:Fn0}) becomes
\bea
F_l=\sum_{m=1}^{N}\gamma_l \gamma_m \int_{-\infty}^{\infty} d\omega | [G(\omega)]_{l,m} |^2
\frac{\omega^2}{\pi}  (T_l-T_m).
\label{eq:FnC}
\eea
Eq. (\ref{eq:FnLR}) and Eq. (\ref{eq:FnC}) are both linear in the SC
baths temperatures. Therefore, we can organize these equations as
$F_l=\sum_{m}C_{l,m}(T_l-T_m)$, with $C_{l,m}$ containing the
frequency integration. Demanding that $F_l=0$ for $l$=2,..,$N-1$, we
get the exact solution \cite{SegalSC}
\bea
{\bf T}=A^{-1} v.
\eea
Here $A$ is a diagonal matrix with $N-2$ rows for $l=2,3,..,N-1$. Its diagonal elements
are $\sum_{m\neq l} C_{l,m}$ and the nondiagonal elements are given
by $-C_{l,m}$. $v$ is a vector defined as
$v_{l}=C_{l,1}T_1+C_{l,N}T_N$. The vector ${\bf T}$ includes the
sought after inner temperatures $T_2$ to $T_{N-1}$. Lastly, given the
vector ${\bf T}$ the current $F_1$ can be readily calculated.


\begin{figure}[htbp]
\vspace{2mm}
{\hbox{\epsfxsize=80mm \epsffile{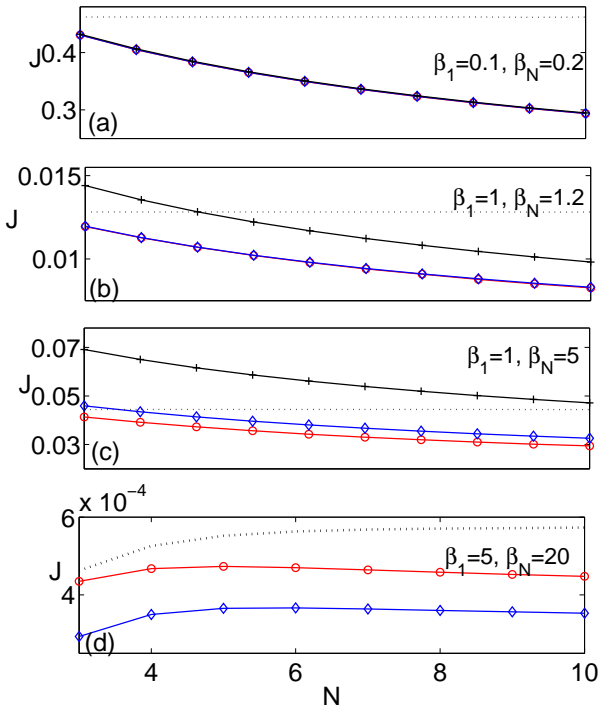}}}
\vspace{3mm}
\caption{Heat current as a function of chain size using the quantum-exact method
($\circ$), quantum linear response formalism (diagonal), and the classical limit ($+$).
$\gamma_{l}=0.2$ for all sites.
(a) $\beta_1=0.1$, $\beta_N=0.2$,
(b) $\beta_1=1$, $\beta_N=1.2$,
(c) $\beta_1=1$, $\beta_N=5$,
(d) $\beta_1=5$, $\beta_N=20$.
The dotted lines represent the exact quantum behavior when the internal reservoirs
are detached from the chain.
}
\label{Fig1}
\end{figure}



\section{Results}

\subsection{Quantum effects in thermal conduction}

We recall that the self consistent reservoirs were introduced as a
tool to include in an effective way anharmonic processes. Our
objective here is to demonstrate novelty in transport mechanisms in
the deep quantum domain, beyond linear response, as a result of the
introduction of these SC reservoirs. Furthermore, we compare
simulations using the classical, quantum linear response, and
quantum-exact treatments, unveiling the importance of quantum
effects at low temperatures, for systems far from equilibrium.
In the simulations reported here quantum results were obtained using
the scheme described in Sec. II.A. We refer to these calculations as
"quantum exact" (QE). We used equation (\ref{eq:FnLR}) to obtain
data in the quantum domain under the linear response approximation,
denoted by "quantum linear-response" (QLR). The classical (C)
behavior was acquired using Eq. (\ref{eq:FnC}).
The following parameters were typically used: unit masses and unit
force constants [see Eq. (\ref{eq:H})], inverse temperatures
$\beta\equiv 1/T$ ranging between $\beta\sim 0.1-20$. Within these
parameters one expects to observe an effective classical behavior
when
$\beta \lesssim 0.2$. We also denote the average temperature by $T_a=(T_1+T_N)/2$
and the overall bias by $\Delta T=T_1-T_N$.

Figure \ref{Fig1} displays the heat current as a function
of size for a uniform chain, comparing data attained from the three
different methods: QE, QLR and C. We analyze the behavior in four
cases: (a) at high temperatures corresponding to the classical
limit, (b) for intermediate temperatures $T_a\sim \omega_S$ and at
small temperature bias $\Delta T<T_a$, corresponding to the
linear-response regime ($\omega_S$ is a characteristic system
frequency), (c) at low temperatures and for far-from equilibrium
situations, $\Delta T/T_a\sim 1$ and, (d) at very low temperatures,
in the deep quantum regime $T_a\ll \omega_S$ and at large bias
$\Delta T/T_a\sim 1$.

As expected, in the high temperature regime the three methods yield the same
value (a). At lower temperatures, yet adopting a small temperature
difference, we find that QE and QLR calculations agree, while
classical simulations overestimate the current (b). At low
temperatures and large bias, panel (c) demonstrates that QLR
calculations overestimate the exact value. In order to appreciate
the role of the SC reservoirs, the dotted line further marks the value of
the heat current, which is obtained within a full quantum
calculations while nullifying the SC reservoirs. We find that in the
absence of these reservoirs the current remains fixed. This is
indeed the expected behavior for harmonic chains with a resonance
thermal conduction mechanism, applicable at high temperatures
\cite{heatcond}.

\begin{figure}[htbp]
\vspace{2mm}
{\hbox{\epsfxsize=80mm \epsffile{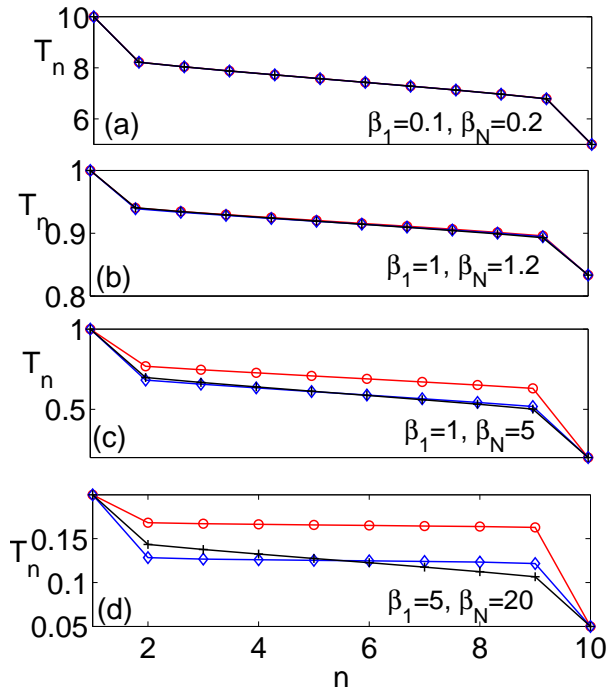}}}
\vspace{3mm}
\caption{Temperature profile of the SC reservoirs at site $n$, $N=10$, using
the exact quantum method ($\circ$), quantum linear response
formalism (diagonal), and the classical expression ($+$),
$\gamma_{l}=0.2$ for all sites.
(a) $\beta_1=0.1$, $\beta_N=0.2$,
(b) $\beta_1=1$, $\beta_N=1.2$,
(c) $\beta_1=1$, $\beta_N=5$,
(d) $\beta_1=5$, $\beta_N=20$.
}
\label{Fig3}
\end{figure}

Fig. \ref{Fig1}(d) exemplifies the heat current behavior at
extremely low temperatures and for a large temperature bias, $\Delta
T/T_a\sim1$. Classical results (not shown) are higher by an order of
magnitude than quantum data. The following observations can be made:
(i) Within the QLR and QE methods, the current demonstrates an
enhancement with size up to $N\sim5$, followed by a decay. However,
{\it in the absence} of the SC internal reservoirs (dotted line) the
current systematically increases with size. This behavior could be
reasoned as follows: With increasing chain length the low frequency
modes of a periodic linear chain are down-shifted \cite{heatcond},
eventually coming into resonance with the populated bath modes. This
effect leads to the enhancement of the heat current with size, the
behavior indeed detected in the absence of the SC baths
\cite{heatcond}. However, when the internal SC reservoirs are added,
scattering mechanisms are responsible for an effective diffusional
motion, resulting in the decay of the current with size
\cite{Dhar-rev}. The combination of these two trends produces the
turnover of the current around $N\sim 5$. (ii) In the deep quantum
regime the QE current is higher that the QLR result, in contrast to
the behavior observed in Fig. \ref{Fig1}(c). As we show in Fig.
\ref{Fig3}(d) below, QLR calculations underestimate the factual
temperature profile, in this case by about $30\%$.  This fact can
explain the similar deviation in the current. On the other hand, QLR
calculations may produce a more significant temperature gradient
within the chain, see e.g.  Fig. \ref{Fig3}(c), resulting in
currents larger than the QE data. As a result of these two
counteracting factors, it is not trivial to predict ad-hoc whether
the accurate quantum result is above or below the QLR limit.

We now display the temperature profile for a chain with $N$=10
particles. We calculate it using the three methods, QE, QLR and C,
in the four parameter domains mentioned above. Fig. \ref{Fig3} shows
that the three calculations generally agree at high temperatures and
for $\Delta T/T_a\ll 1$. In contrast, at very low temperatures and
for $\Delta T/T_a\sim 1$ the QE results deviate from the QLR and the
classical data in a profound way. Specifically, for this (symmetric)
setup QLR and classical calculations provide reservoirs temperatures
which {\it symmetrically} vary around the average temperature
$T_a=(T_1+T_N)/2$ \cite{SegalSC}. In contrast, QE calculations
reveal that the chain temperature is actually {\it higher} than this
temperature. This shifted profile stems from the nonlinear
Bose-Einstein distribution function characterizing the reservoirs
statistics.
It is also of interest to note that classical and QLR simulations generally
predict an internal temperature gradient which is higher than the QE value.
Specifically, with classical simulations we compute a local gradient which is larger by an order of
magnitude from the QLR and the QE behavior [Fig. \ref{Fig3}(d)].
This considerable disagreement reflects itself in the classical heat current, in this case
high by one order of magnitude compared to the exact results.


\begin{figure}[htbp]
\vspace{2mm}
\hspace{2mm}
{\hbox{\epsfxsize=75mm \epsffile{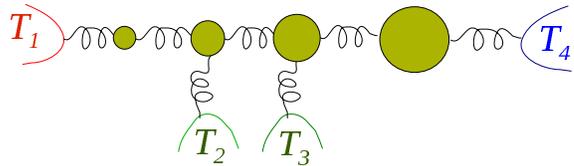}}}
\vspace{3mm}
\caption{A scheme of the mass-graded harmonic chain with SC baths for $N=4$ beads.}
\label{FigMR}
\end{figure}


\subsection{Application: Quantum thermal rectification}

Thermal rectification, an asymmetry of the heat current for forward
and reversed temperature gradients, has been extensively analyzed in
the last decade \cite{Rectifier,RectifierE}. In a desirable
rectifier the system behaves as an excellent heat conductor in one
direction of the temperature bias, while for the opposite direction
it effectively acts as an insulator. It is agreed that junctions
incorporating anharmonic interactions with some sort of spatial
asymmetry should exhibit this effect \cite{Rectifier}. Since the SC
model includes, in an effective way, anharmonic interactions through
the action of the SC reservoirs, it is of interest to explore
whether this model could demonstrate the rectifying effect when some
spatial asymmetry is incorporated. This question is of particular
interest since neither the classical SC model nor the QLR SC case
can show thermal rectification, even when asymmetry is introduced
\cite{Pereira, SegalSC,PereiraQ}. In a recent paper analytical arguments were
put forward, indicating that quantum SC systems should rectify
heat \cite{PereiraPLA}. In what follows we demonstrate that this
effect indeed exists in asymmetric quantum harmonic systems with SC
baths. We incorporate asymmetry either by using a mass-graded chain,
see Fig. \ref{FigMR}, or by connecting the chain asymmetrically
to the reservoirs at the boundaries.

\begin{figure}[htbp]
\vspace{2mm}
{\hbox{\epsfxsize=80mm \epsffile{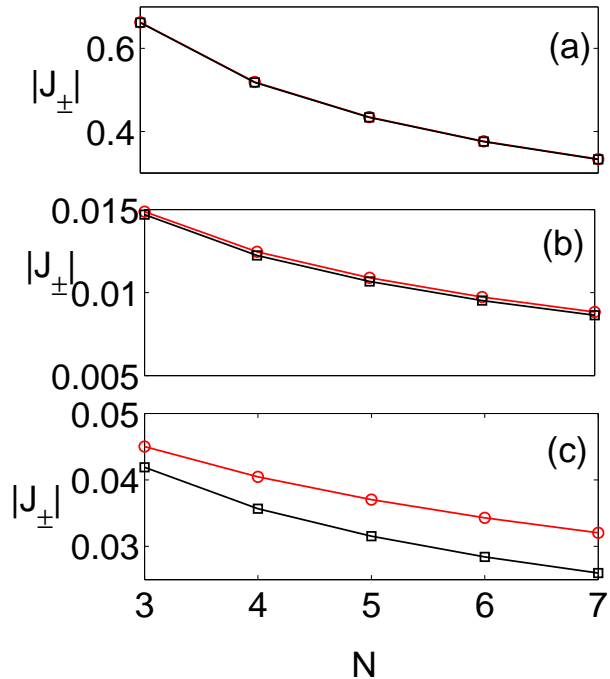}}}
\vspace{3mm}
\caption{Thermal rectification in a mass graded system:
The magnitude of the heat current as a function of chain size
for forward temperature bias $T_1>T_N$ ($\circ$) and for the backward direction,
$T_N>T_1$ (square), $\gamma_{n}=0.2$ for all sites, $M_1=0.2$, $M_n=M_1+(n-1)\times0.2$.
In the forward direction we used
(a) $\beta_1=0.1$, $\beta_N=0.2$,
(b) $\beta_1=1$, $\beta_N=1.2$, and
(c) $\beta_1=1$, $\beta_N=5$.
 The opposite polarities were used in each case to generate $J_-$.}
\label{Fig4}
\end{figure}

Fig. \ref{Fig4} displays the absolute value of the heat current for
forward ($J_+$) and reversed ($J_-$) temperature biases, studying a
mass-graded system with $M_1=0.2$ and $M_l=M_1+0.2\times (l-1)$,
using the QE method. While at high temperatures and for $\Delta T
\ll T_a$ the effect is negligible and $J_+\sim J_-$, at low
temperatures and for large bias $J_+$ and $J_-$ evidently deviate,
with the current being larger in the direction of {\it increasing}
masses. We also find that the rectification ratio $|J_+/J_-|$ is
increasing with chain size, an observation which can be reasoned by
the growing mass difference along the chain.


We note that in  different experimental and theoretical studies,
e.g., Refs. \cite{RectifierE,massRev,mass1,mass2}, the opposite
tendency has been reported, and the preferred direction of heat
transfer occurs from heavy to light atoms. It is clear that the
preferred direction depends on the details of the system studied
\cite{massRev,mass1,mass2,mass5}. For example, Ref. \cite{mass1}
reports on molecular dynamics simulations of heat conduction in
mass-graded chains assuming anharmonic interactions between
particles. The preferred heat transport direction in that model
(heavy to light) is attributed to the vibrational coupling between
low and high modes, arising due to anharmonicity in the system. The
dynamics is further interpreted in terms of the overlap between the
power spectra at the chain ends. In contrast, simulations of thermal
conduction in mass-graded nanotubes showed the opposite trend,
explained by the transfer of vibrational energy from the transverse
to the longitudinal direction \cite{mass3,mass4}.

\begin{figure}[htbp]
\vspace{2mm}
{\hbox{\epsfxsize=80mm \epsffile{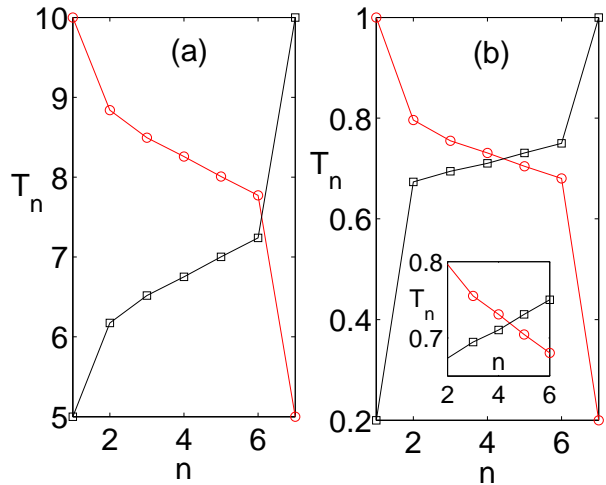}}} \vspace{3mm}
\caption{Temperature profile for a mass-graded $N=7$ chain. In the forward
direction ($\circ$) we used (a) $\beta_1=0.1$, $\beta_N=0.2$ and (b)
$\beta_1=1$, $\beta_N=5$. The opposite polarities were used to
generate the reversed profile (square). Other parameters are the
same as in Fig. \ref{Fig4}.
The inset zooms on the chain center.}
\label{Fig4T}
\end{figure}

In order to better understand the mechanism of thermal rectification
in our model we display the temperature profile for the rectifying
system in Fig. \ref{Fig4T}. At high temperatures (a) there is a
reflection symmetry with respect to the average temperature, when the
temperature bias is reversed \cite{SegalSC}. In contrast, in the quantum domain
beyond linear response an asymmetry is discovered (b): 
The {\it temperature gradient} at the chain center is larger when the 
light masses are in contact with the hot bath, than the gradient generated in the reversed case.
In particular, $\nabla T\sim -0.0230$ (-0.0185) when the
heat flows in the direction of increasing (decreasing) masses
weight. The ratio between these gradients indeed fits $|J_+/J_-|$ for $N=7$, see Fig. \ref{Fig4}(c).
We emphasize that arguments based on the power spectra overlap
\cite{mass1, mass2} cannot be put on for the harmonic SC model since
it does not include a physical mechanism for coupling different vibrational
modes. This can be seen in Eq. (\ref{eq:Fn0}): The heat current is
retrieved by integrating over {\it separate} contributions, summing
different frequency components.

\begin{figure}[htbp]
\vspace{2mm} {\hbox{\epsfxsize=80mm \epsffile{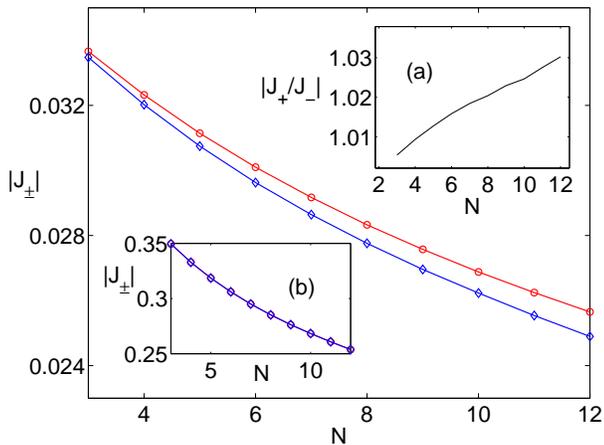}}}
\vspace{3mm} \caption{Thermal rectification in systems with contact
asymmetry: The magnitude of the heat current as a function of chain
size for forward temperate bias $T_1>T_N$ ($\circ$) and for the
backward direction, $T_N>T_1$ (square) $\gamma_{1}=0.1$,
$\gamma_{N}=0.4$, $\gamma_{n\neq 1,N}=0.2$. $\beta_1=1$,
$\beta_N=5$, and the the reversed contact symmetry. (a) Rectification ratio
$|J_+/J_-|$. (b) The forward and backward currents at high
temperature $\beta_1=0.1$ and $\beta_N=0.2$, and the reversed setup,
with no significant rectification observed. } \label{Fig5}
\end{figure}

One could also generate a spatial asymmetry in a mass-uniform SC model
by coupling it unequally  to the two ends,  using
$\gamma_1\neq \gamma_N$. Since this is a contact asymmetry, one
would generally expect its effect to diminish with size. Fig.
\ref{Fig5} still shows a small enhancement of the rectification
strength with $N$, probably due to the increased importance of the
scattering mechanisms (mimicking anharmonicity) with size.

Concluding this section, an interesting outcome of our study is the
confirmation that the quantum harmonic chain with SC baths acts as a
pure quantum thermal rectifier, since its classical analog, or the
quantum model in the linear response regime, can not demonstrate
this effect  \cite{PereiraPLA}. This behavior is attributed to the
introduction of the SC reservoirs, whose statistics at low
temperatures is a {\it nonlinear} function of the local temperature,
unlike the high-temperature or linear response behavior.

\section{Summary}

We developed a numerical method for acquiring the heat current and
the temperature profile in the quantum harmonic chain model with SC
reservoirs, beyond the linear response approximation. While the
technique is generally valid for 3D models, we applied it here on
1D linear chains. At low temperatures and for large temperature
biases we found that the exact quantum results significantly deviate
from the QLR and classical behavior. As an application, we explored
the thermal rectification effect in asymmetric systems, either by
introducing mass asymmetry or by imposing a contact asymmetry. In
both cases we concluded that quantum statistics is responsible for
the onset of the nonlinear rectifying effect.

Our method could be generalized in two nontrivial ways. First, the
scheme could be feasibly extended to describe non-ohmic reservoirs,
for studying the role of memory effects on thermal transport. This
could be done by introducing frequency dependent friction terms in
Eq. (\ref{eq:Fn0}). Since the method is fully numerical, one can
easily incorporate such frequency-dependent friction terms in the
calculations.
The second utility of the method is its application to fermionic
systems. The analogous electronic model has been of extreme interest
\cite{fermion}, where the SC reservoirs are interpreted as local
dephasing probes \cite{Butt}. For quantum systems, results were
obtained only under the linear response assumption
 \cite{scelectron}.
The principle introduced here could be easily modified, for treating
the electronic problem, by replacing the bosonic distribution
functions by fermionic functions, and by revising the equations
as necessary. We expect that quantum effects will play a
significant role at low temperatures and for large potential biases
in the electronic case, similarly to the behavior found in the
present phononic model.


To conclude, our calculations indicate that at low temperatures and
for large biases the thermal conduction of SC quantum harmonic
chains is fundamentally distinct from the linear response behavior
or classical characteristics, manifesting  interesting
functionality that does not take place at high temperatures or close
to equilibrium. It is of interest to extend our simulations and
study longer chains, $N\sim 50$, for understanding the scaling of
the heat current with size in the deep quantum domain and far from
equilibrium.

\begin{acknowledgments}
The work was supported by an NSERC discovery grant.
\end{acknowledgments}


\begin{thebibliography}{99}
\bibitem{leak}
The central reservoirs, if not optimally tuned as SC baths, may
either absorb or donate energy to the chain, see Fig. \ref{Fig0}
showing internal currents with opposite signs. In the text we often
refer to this net exchange of energy as "leakage".
\end{thebibliography}
\end{document}